\documentclass{svproc}
\usepackage{xcolor,float} 

\usepackage{url,graphicx, multicol,footmisc,amsmath,amssymb,epstopdf}
\usepackage{multirow,booktabs,makecell,subcaption}

\begin{document}
\mainmatter              
\title{A MgNO Method for Multiphase Flow in Porous Media}
\titlerunning{MgNO for Porous Media Flow}  
%
\author{Xinliang Liu\inst{1} \and Xia Yang\inst{2} \and Chen-Song Zhang\inst{3}\thanks{Corresponding author} \and Lian Zhang\inst{4} \and Li Zhao\inst{3}}

\authorrunning{Xinliang Liu et al.} 
%
\tocauthor{Xinliang Liu, Xia Yang, Chen-Song Zhang, Lian Zhang, Li Zhao}
\institute{
Computer, Electrical and Mathematical Science and Engineering Division, King Abdullah University of Science and Technology, Thuwal, Saudi Arabia
\and
Xiangtan University, Hunan, China
\and
LSEC, ICMSEC, Academy of Mathematics and System Sciences, Chinese Academy of Sciences \& University of Chinese Academy of Sciences, China
\and
Shenzhen International Center for Industrial and Applied Mathematics, Shenzhen Research Institute of Big Data
}

\maketitle              

\begin{abstract}
This research investigates the application of Multigrid Neural Operator (MgNO), a neural operator architecture inspired by multigrid methods, in the simulation for multiphase flow within porous media. The architecture is adjusted to manage a variety of crucial factors, such as permeability and porosity heterogeneity. The study extendes MgNO to time-dependent porous media flow problems and validate its accuracy in predicting essential aspects of multiphase flows. Furthermore, the research provides a detailed comparison between MgNO and Fourier Neural Opeartor (FNO), which is one of the most popular neural operator methods, on their performance regarding prediction error accumulation over time. This aspect provides valuable insights into the models' long-term predictive stability and reliability. The study demonstrates MgNO's capability to effectively simulate multiphase flow problems, offering considerable time savings compared to traditional simulation methods, marking an advancement in integrating data-driven methodologies in geoscience applications.

\end{abstract}
\section{Introduction}
In recent years, the application of deep learning techniques in scientific computing has shown promising potential to revolutionize traditional simulation methods. This paper presents the development of a deep learning-enabled simulation method for porous media, aimed at addressing the challenges in modeling and simulating complex multiphase multicomponent flow in porous media. 

The simulation framework, named as OpenCAEPoro (\url{https://github.com/OpenCAEPlus/OpenCAEPoroX}), is based on a set of partial differential equations (PDEs) for multicomponent multiphase porous media flow. The Darcy's law is employed to describe the volumetric flow rate of each phase in multiphase flow, which depends on factors like pressure differences, rock permeability, and phase viscosity. The model also incorporates phase balance, saturation, and equations of state (EOS) to provide a comprehensive description of the physical phenomena involved. The OpenCAEPoro simulator has been extensively tested with the standard benchmark problems from the SPE (Society of Petroleum Engineers) comparison projects. These benchmarks are critical in validating the accuracy and performance of the simulation framework compared to established commercial software solutions. The benchmark tests have demonstrated the capability and parallel scalability of the framework to handle complex reservoir simulations effectively. 

The integration of deep learning techniques with advanced parallel computing methods offers a powerful approach to simulating complex porous media flows. The developed framework demonstrates significant potential in improving simulation accuracy, efficiency, and scalability, paving the way for more robust applications in reservoir engineering and beyond.
When addressing multiphase flow problems, deep learning methods can directly learn flow characteristics and patterns from fluid data without the need for understanding the underlying physical mechanisms. This data-driven approach is particularly suitable for handling highly nonlinear and multiscale interactions in flow problems. Even in situations where data is scarce or conditions change, the models can maintain reasonable performance. Moreover, the rapid prediction feedback of deep learning models provides a powerful tool for real-time flow analysis, history matching, and inverse problems in reservoir simulation, as well as offering significant applications in engineering optimization and disaster warning.

We investigate an emerging deep learning method for solving multiphase flow within porous media problems — the Multigrid Neural Operator (MgNO) model~\cite{he2024}. A series of numerical experiments are designed to provide substantial scientific evidences for its promotion in practical applications. The main contributions of this paper are listed as follows:
\begin{itemize}
    \item \textbf{Introducing time variable to the model:} The input features include a temporal channel, which captures the dynamic behavior of the PDE system over time and makes predicting time-dependent phenomena possible. The temporal channel enables the MgNO model to handle dynamic flow problems, enhancing its adaptability to various application scenarios.
    
    \item \textbf{Performance evaluation and comparison:} The proposed model is compared with other existing deep learning models such as Fourier Neural Operator (FNO). The effects of different loss functions on model training are evaluated. The numerical results verify MgNO's performance in multiphase flow problems, and reveal its performance advantages and applicable range.
    
    \item \textbf{Application exploration and generalization tests:} Generalization abilities of MgNO and FNO for time evolution problems are tested. MgNO demonstrates superior generalization performance for predicting unseen temporal features in practice.
\end{itemize}

\section{Literature Review}
\subsection{Neural Networks Method}
Neural networks (NN) are artificial intelligence models inspired by biological neural systems, composed of multiple layers of neurons. Each layer is connected to adjacent layers, and the network learns by adjusting connection weights to recognize and learn complex patterns.

In reservoir models, various deep learning methods have been successfully applied and explored, with different techniques used to solve various problems and tasks \cite{hemmati2020}. Using CNNs to process seismic data to identify subsurface structures and guide reservoir exploration \cite{geng2020}. Additionally, CNNs can analyze core images to identify rock types and porosity, supporting reservoir prediction \cite{cheng2017}. LSTM-based neural networks model reservoir production data to predict well output and pressure changes \cite{song2020}. A GAN-based geological model generation method generates realistic geological models to assist reservoir exploration and development decisions \cite{zhang2019}. Using DDPG to learn optimal water-flooding schemes for maximizing recovery rates and economic benefits \cite{ma2019}. Introducing deep learning and particle swarm optimization (PSO) techniques to predict permeability, guiding the construction of the reservoir pore-throat system and evaluating reservoir flow capacity \cite{gu2019}. Applying auto-encoders for geological data anomaly detection to identify abnormal conditions in formations \cite{chow2020}. Feng et al. \cite{feng2024encoder} uses the fusion of CNN and LSTM (ConvLSTM) as the surrogate model for simulating geological $\text{CO}_2$ sequestration.

\subsection{Neural Operators Learning}
Traditional machine learning models learn the mapping relationship between inputs and outputs, but operator learning extends to learning and approximating the entire function space, enabling it to handle more complex problems such as solving differential equations and modeling dynamic systems. The key to operator learning is to treat input and output as elements in function spaces, learning a mapping between input function space and output function space, thereby approximating the operator. This method can effectively handle problems involving complex operations such as differential and integral operators, widely applied in solving differential equations, optimization problems, and more.

Operator learning methods can effectively learn and approximate the nonlinear characteristics of differential operators, achieving efficient solutions for partial differential equations [17]. An improved MgNet model in \cite{he2019} solves numerical partial differential equations through operator learning [18].

Utilizing the powerful fitting capabilities of neural networks, neural operators can learn nonlinear relationships in data to approximate complex operators or functions. Neural networks, through multiple layers of nonlinear transformations and activation functions, can approximate arbitrarily complex nonlinear functions. In reservoir simulation, neural operators can learn nonlinear relationships in data such as permeability, relative permeability, and saturation, achieving accurate modeling and precise solutions for multiphase flow problems. Their applications involve permeability estimation, relative permeability prediction, and more.

The FNO in \cite{li2020} and MgNO in \cite{he2024} models are two advanced deep learning tools that have shown remarkable advantages in the field of multiphase flow. FNO, by learning the solutions of partial differential equations in the frequency domain, effectively captures the wave properties and nonlinear characteristics of multiphase flow. MgNO combines deep learning with the multigrid method, further enhancing the efficiency and accuracy of solving multiphase flow problems. These models demonstrate the tremendous potential of deep learning in addressing multiphase flow issues.

\subsubsection{FNO Method Overview.}
The core idea of FNO is to map input data space to frequency space through Fourier transform and then use neural networks to learn this mapping process. The advantage of Fourier transform in handling wave problems enables FNO to efficiently handle a wide range of PDE problems.

\begin{figure}[h]
    \centering
    \includegraphics[width=0.8\textwidth]{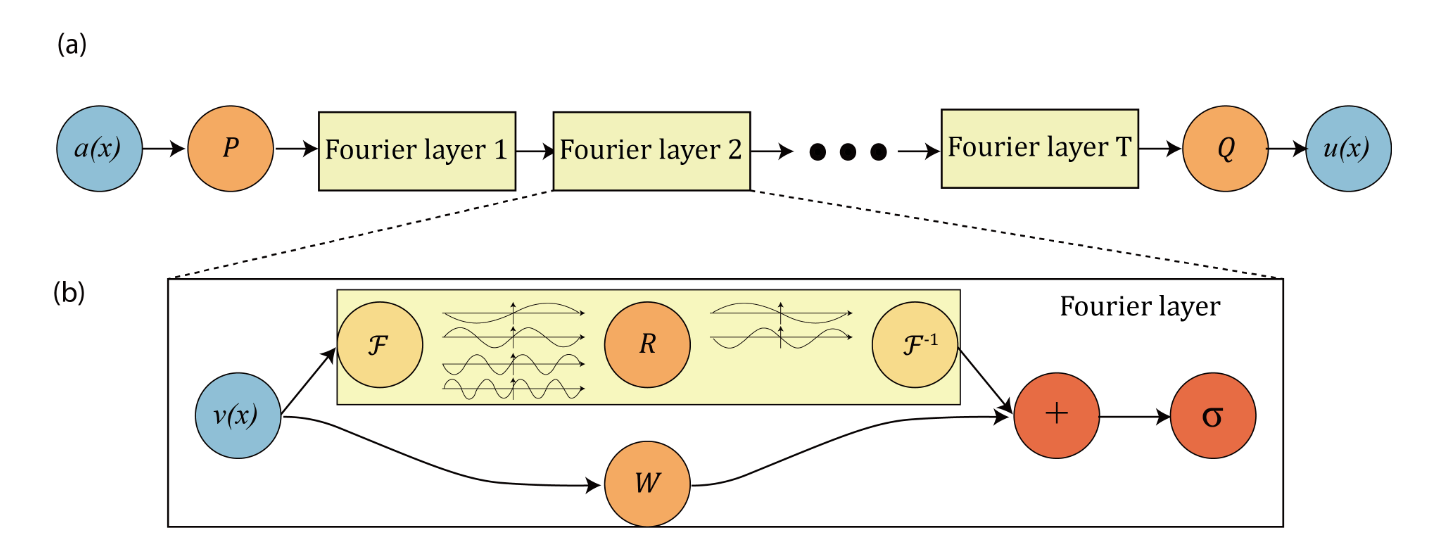}
    \caption{\textbf{top:} The architecture of the FNO; \textbf{bottom:} Fourier layer.}
    \label{fig:FNO Model}
\end{figure}

FNO learns the mapping from one infinite-dimensional function space to another. Unlike traditional solution methods, neural operators can directly learn the solution mappings of parameterized PDEs, learning the solutions of an entire family of PDEs rather than a single instance. FNO achieves a highly expressive and efficient architecture by directly parameterizing the integral kernel in the Fourier space.The structure of the FNO model mainly includes the following aspects:

\begin{itemize}
    \item \textbf{Global Convolution Operation: }FNO significantly enhances the model's ability to handle large-scale PDE problems through fast Fourier transform (FFT) for global convolution processing of inputs. This global convolution operation allows FNO to capture the global information of input data, contrasting sharply with traditional local convolution operations.
    \item \textbf{Parameterization Strategy: }FNO employs linear transformations in the Fourier space to handle low-frequency modes while filtering high-frequency modes. Therefore, FNO can effectively learn dynamic changes in the frequency space, capturing the global nature of PDE solutions.
\end{itemize}

The FNO learns a mapping between infinite-dimensional spaces from observed input-output pairs. The methodology approximates the operator mapping coefficients to solutions of parametric PDEs. Formally, we define the operator \(G\) as follows:
\begin{equation}
G : A \times \Theta \rightarrow U \quad \text{or equivalently} \quad G_{\theta} : A \rightarrow U, \, \theta \in \Theta,
\end{equation}
where \(A\) and \(U\) are separable Banach spaces representing the input and output function spaces, respectively. The goal is to learn the parameters \(\theta \in \Theta\) such that the neural operator \(G_{\theta}\) approximates the true operator \(G^{\dagger}\).

The architecture iteratively updates representations using a combination of non-local integral operators and local nonlinear activation functions. The iterative process involves the following steps:

\begin{itemize}
    \item \textbf{Lifting}: The input function \(a(x)\) is lifted to a higher-dimensional representation \(v_0(x) = P(a(x))\) using a neural network \(P\).
    \item \textbf{Iterative Updates}: The representation is updated iteratively through a sequence of layers, each consisting of a non-local integral operator \(K\) and a local, nonlinear activation function \(\sigma\):
    \begin{equation}
    v_{t+1}(x) := \sigma \left( W v_t(x) + \left( K(a; \phi) v_t \right)(x) \right), \quad \forall x \in D,
    \end{equation}
    where \(W\) is a linear transformation, and \(K(a; \phi)\) is a kernel integral operator parameterized by \(\phi\).
    \(K\) is defined as:
    \begin{equation}
    (K(a; \phi) v_t)(x) = \int_D \kappa(x, y, a(x), a(y); \phi) v_t(y) \, dy.
    \end{equation}
    Here, \(\kappa_\phi : \mathbb{R}^{2(d+d_a)} \rightarrow \mathbb{R}^{d_v \times d_v}\) is a neural network parameterized by \(\phi \in \Theta_K\). Even though the integral operator is linear, the neural operator can learn highly non-linear operators by composing linear integral operators with non-linear activation functions, analogous to standard neural networks.
    \item \textbf{Projection}: The final output \(u(x)\) is obtained by projecting the high-dimensional representation back to the target dimension using another neural network \(Q\): 
    \begin{equation}
     u(x) = Q(v_T(x)).   
    \end{equation}
\end{itemize}

The integral operator in the update rule is replaced by a convolution operator defined in Fourier space, leveraging the Fast Fourier Transform (FFT) for efficient computation. Specifically, the kernel integral operator \(K\) is parameterized in Fourier space as follows:
\begin{equation}
\left( K(\phi) v_t \right)(x) = \mathcal{F}^{-1} \left( R_{\phi} \cdot \mathcal{F}(v_t) \right)(x), \quad \forall x \in D.
\end{equation}
Here, \(\mathcal{F}\) and \(\mathcal{F}^{-1}\) denote the Fourier transform and its inverse, respectively. \(R_{\phi}\) is a linear transformation applied in the Fourier domain. This parameterization allows the method to efficiently handle high-dimensional data and achieve state-of-the-art performance in solving PDEs.

The core advantage of FNO lies in its resolution invariance of solutions. It can be trained at low resolution and perform zero-shot super-resolution at higher resolutions without retraining the model. This capability is particularly important for complex fluid dynamics problems, as it allows the model to capture small-scale phenomena such as vortices often appearing in high-resolution simulations.

Furthermore, FNO achieved resolution-invariant solution operators in turbulent modes for the first time when handling Navier-Stokes equations, showing better convergence than previous graph-based methods. FNO not only demonstrated excellent performance at fixed resolutions but also maintained unchanged network parameters when increasing the resolution of input and output spaces, showing high flexibility and adaptability.

\subsubsection{MgNO Method Overview.}
Traditional multigrid methods solve problems at different accuracy levels to achieve fast convergence. MgNO borrows this idea, simulating solution spaces at multiple accuracy levels to achieve efficient PDE solutions. This method effectively utilizes the advantages of multigrid methods in handling multiscale data, significantly reducing model training time and improving solution accuracy.

MgNO defines neural operators as connections in neural networks, where each connection is regarded as a bounded linear operator, contrasting with traditional neural network connections. This definition eliminates the need for lifting and projecting operators commonly required in traditional neural operators and can naturally adapt to various boundary conditions. Moreover, compared to other CNN-based models, MgNO demonstrates better performance in training and generalization capabilities, especially when handling PDE problems with complex boundary conditions.
\begin{figure}[h]
    \centering
    \includegraphics[width=0.8\textwidth]{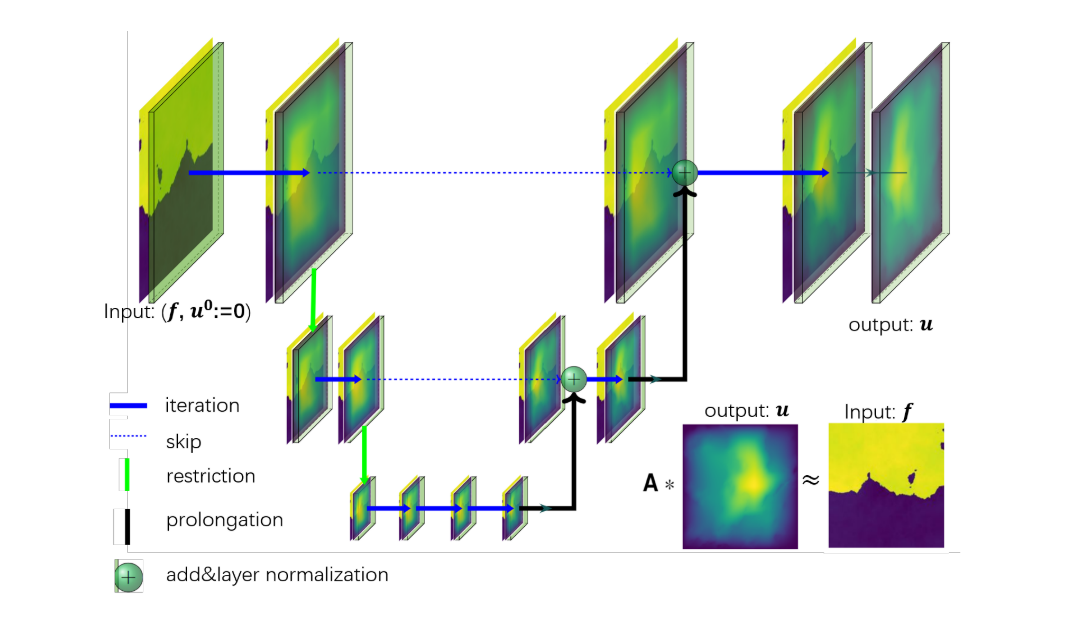}
    \caption{MgNO Network Architecture}
    \label{fig:FNO Model}
\end{figure}
The neural operator is defined analogously to a fully connected neural network. The output \( O_i(u) \) of the i-th neuron in a nonlinear operator layer is:
\begin{equation}
    O_i(u) = \sigma \left( \sum_j W_{ij} u + B_{ij} \right),
\end{equation}
where \( W_{ij} \) denotes the bounded linear operator and \( B_{ij} \) is a bias function.

A universal approximation theorem is introduced for neural operators, ensuring that any continuous operator can be approximated by the proposed neural network structure:
\begin{equation}
    \inf_{O \in \Xi_n} \sup_{u \in C} \|O^*(u) - O(u)\|_Y \leq \epsilon,
\end{equation}
where \( X = H^s(\Omega) \) and \( Y = H^{s'}(\Omega) \) are Banach function spaces, \( O^* \) is the target operator, \( O \) is the neural operator, \( C \subset X \) is a compact set, \( \epsilon \) is an arbitrarily small positive number, and \( \Xi_n \) denotes the set of shallow networks defined with \( n \) neurons.

The multigrid method is used to parameterize the linear operators \( W_{ij} \) in the neural network. This method is efficient in handling various boundary conditions and offers high expressivity with minimal parameters. The V-cycle multigrid process for solving an elliptic PDE is represented as a convolution neural network with one channel:
\begin{equation}
    \|u - W_{Mg}(A * u)\|_A \leq \left(1 - \frac{1}{c}\right) \|u\|_A,
\end{equation}
where \( c \) is a constant independent of the mesh size \( h \), and \( \|u\|_A = (u, A * u)_{L^2(\Omega)} \) denotes the energy norm.

The shallow neural operator with \( n \) neurons is defined as:
\begin{equation}
    O(u) = \sum_{i=1}^n A_i \sigma(W_i u + B_i),
\end{equation}
where \( W_i \in L(X, Y) \), \( B_i \in Y \), and \( A_i \in L(Y, Y) \).

The deep neural operator with \( L \) hidden layers and \( n_{\ell} \) neurons in the \( \ell \)-th layer is defined as:
\begin{equation}
    \begin{cases}
h_0 = u \in X ,\\
h_{\ell}(u) = \sigma \left( W_{\ell} h_{\ell-1}(u) + B_{\ell} \right) \in Y^{n_{\ell}}, & \ell = 1 : L, \\
O(u) = W_{L+1} h_L(u) \in Y.
    \end{cases}
\end{equation}

The article explains how multigrid methods are used to parameterize the operator \( W_{\ell} \) by representing it with multi-channel convolutions. This is done by extending the multigrid approach to handle multi-channel inputs and outputs.

The MgNO operator maps from the linear finite element space of input functions to the linear finite element space of output functions:
\begin{equation}
    \begin{cases}
    h_0 = u \in X, \\
    h_{\ell}(u) = \sigma(W_{\ell}^{Mg} h_{\ell-1}(u) + B_{\ell} h_{\ell-1}(u) + b_{\ell} 1) \in Y^n, & \ell = 1 : L \\
    v = \tilde{G}_{\theta}(u) = W_{L+1}(h_L(u)) \in Y.
    \end{cases}.
\end{equation}
where \( W_{\ell}^{Mg} \) represents the multi-channel V-cycle multigrid operator. Unlike UNet in \cite{ronneberger2015}, which utilizes a single V-cycle, MgNO employs multiple V-cycles, with each linear operator acting as a V-cycle, thereby distinguishing it from MgNet in \cite{he2019}.

MgNO embeds a multigrid structure into deep neural networks, optimizing the parameterization process of the network. This model has the following characteristics:
\begin{enumerate}
    \item \textbf{Linear Operator Parameterization:}MgNO utilizes multigrid strategies to effectively parameterize linear operators in neural networks, simplifying network structure and improving computational efficiency.
    \item \textbf{Boundary Condition Handling: }MgNO can naturally handle various boundary conditions, thanks to its adaptive design in the parameterization process.
    \item \textbf{Training and Generalization Capabilities:}MgNO exhibits easier training and better generalization capabilities, especially in handling PDE problems with complex boundary conditions.
\end{enumerate}
By combining the advantages of multigrid methods and neural networks, the MgNO model provides an efficient and accurate PDE solution framework. This model not only has a solid theoretical foundation but also shows significant performance advantages in practical applications.

\section{Mathematical Model}

We consider the mass balance equation for multiphase flow
\begin{equation}
    \frac{\partial}{\partial t}\left(\phi \sum_\alpha S_\alpha \rho_\alpha \right)-\nabla \cdot\left\{K \sum_\alpha \frac{k_{r \alpha}}{\mu_\alpha} \rho_\alpha  \nabla p \right\}+ q_\alpha =0,
\end{equation}
where the first term is the accumulation term for fluid storage within the rock pore volume; the second is the flux; the third is the source or sink term. The subscript $\alpha$ denotes the fluid phase; $t$ is time; $\phi$ is the rock porosity, $S_\alpha$ is the phase saturation; $\rho_\alpha$ is the fluid phase density; $K$ the rock permeability; $k_{r \alpha}$ is the relative permeability of phase $\alpha ; \mu_\alpha$ is the phase viscosity; $p_\alpha$ is the phase pressure;  and $q_\alpha$ denotes the rate for extracting or injecting fluid phase $\alpha$.

Consider the oil-water two-phase model with a grid size of 128×1×128 (xoz plane), uniformly divided in the x, y, and z directions with a step size of 10. The initial porosity is 0.3, and the initial water saturation is 0.2. There is an injection well on the left boundary and a production well on the right boundary. The simulation lasts for 24 days, with results output every day.

Additionally, several auxiliary relationships constrain the variables in the mass balance equation. The pore space is filled by both fluid phases, 
$
\sum S_\alpha=1.
$
The distribution of absolute permeability \(K \sim  N(0, A(-\Delta + 9I)^{-2})\), where \(A = 10\) is the jump amplitude. \(K\) is independent of time, and if written as a time series, it can be expressed as \(K(x, t_0) = K(x, t_1) = \cdots = K(x, t_{24}) = K(x)\); since \(K \geq 0\), the absolute value is added. The larger the \(K\), the faster the fluid flows, so \(A\) is used to adjust the range of \(K\). In each cell, \(K_i\) is a vector (let the absolute permeabilities in the x, y, and z directions be \(K_{x_i}, K_{y_i}, K_{z_i}\)), considering isotropic permeability, i.e., \(K_{x_i} = K_{y_i} = K_{z_i} = K_i\).

\begin{figure}[h]
    \centering
    \includegraphics[width=0.8\textwidth]{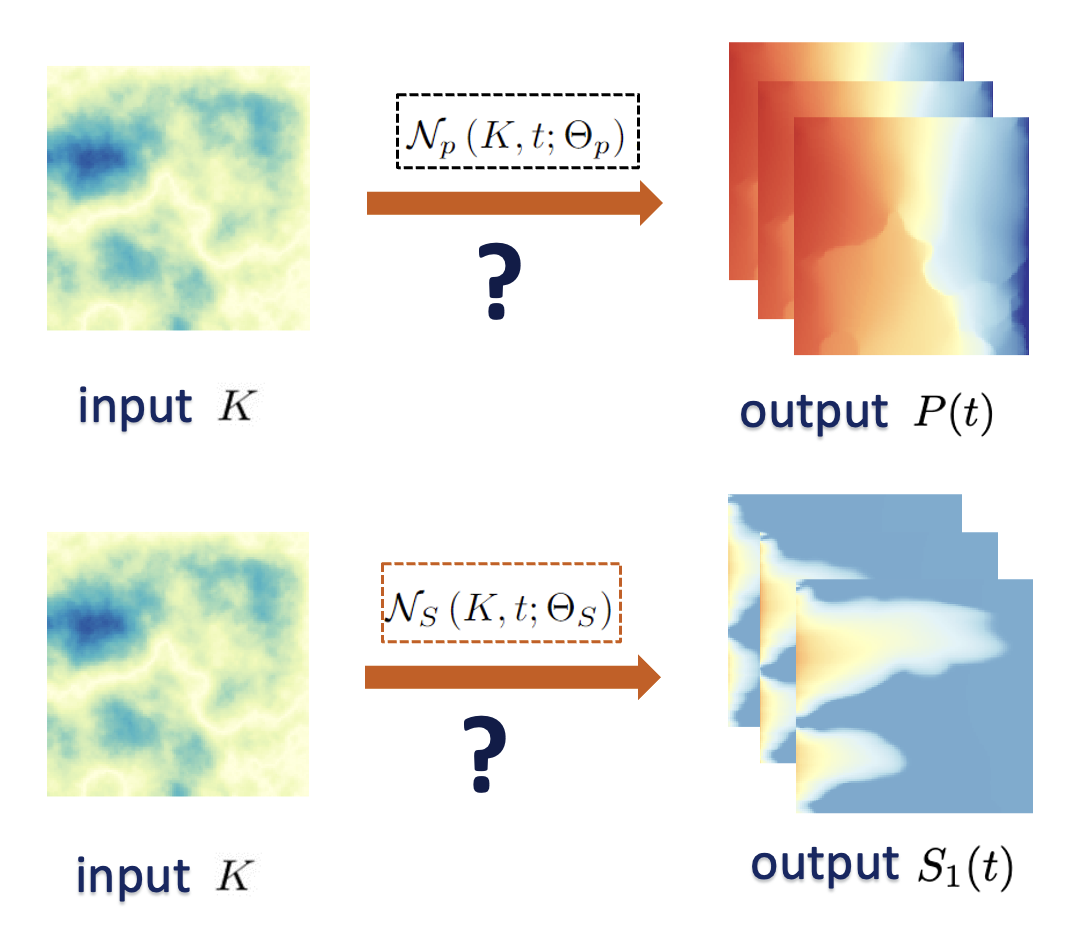}
    \caption{Neural Operator \(N_P\) and \(N_S\)}
    \label{fig:KtoPSw}
\end{figure}

From Figure \ref{fig:KtoPSw}, our research objective is to train two neural operators, \(N_P\) and \(N_S\), to respectively predict the temporal variations of pressure \(P(t)\) and water saturation \(S_1(t)\). The KP model represents using permeability \(K\) to predict the temporal variation of pressure \(P(t)\); the KS model represents using permeability \(K\) to predict the temporal variation of water saturation \(S_1(t)\).

\begin{enumerate}
    \item \textbf{Neural Operator \(N_P\):}
    \begin{itemize}
        \item \textbf{Input:} Rock permeability field \(K\), time \(t\), and parameters \(\theta_P\).
        \item \textbf{Output:} Temporal variation of the pressure field \(P(t)\).
        \item \textbf{Task:} This operator aims to capture the impact of complex geological structures and permeability variations on the evolution of the pressure field. By learning the pressure field distribution at different time steps, it achieves predictions of the dynamic pressure field.
    \end{itemize}
    
    \item \textbf{Neural Operator \(N_S\):}
    \begin{itemize}
        \item \textbf{Input:} Rock permeability field \(K\), time \(t\), and parameters \(\theta_S\).
        \item \textbf{Output:} Temporal variation of the water saturation \(S_1(t)\).
        \item \textbf{Task:} This operator primarily focuses on the evolution of water phase saturation in multiphase flow. By learning the saturation distribution at different time steps, it achieves predictions of the dynamic saturation field.
    \end{itemize}
\end{enumerate}

By introducing the neural operator framework, we aim to accurately capture the flow characteristics under complex geological conditions while maintaining computational efficiency. This provides an efficient and accurate prediction tool for applications in reservoir engineering, groundwater flow, and other related fields.

\section{Experimental Results and Analysis}
\subsection{Model Inference}

We evaluated the trained models using a dataset of 400 samples from time steps 0 to 24 by calculating the relative \( L_2 \) error and the prediction time for each time series sample, as shown in Table \ref{tab:example}.

The MgNO model's larger number of parameters leads to a longer processing time per sample compared to the FNO. However, its lower relative \( L_2 \) error in the more complex K-S task suggests that MgNO's performance advantage may outweigh the longer computation time. Compared to classical simulation, which requires 0.164 seconds per sampling step, deep learning models improve the speed by 125 times.

\begin{table}[h]
    \centering
    \caption{Model Performance Test Results}
    \label{tab:example}
    \begin{tabular}{cccccc}
        \toprule
        \multirow{2}{*}{\textbf{Prediction}} & \multirow{2}{*}{\textbf{Model}} & \multirow{2}{*}{\makecell{\textbf{Parameter}\\\textbf{Count}}} & \multirow{2}{*}{\makecell{\textbf{Relative}\\\textbf{L2 Error}}} & \multirow{2}{*}{\makecell{\textbf{Time (s)}\\\textbf{per Sample}}} \\ 
        & & & & \\ \midrule
        \multirow{4}{*}{K-P} & MgNO-\( L_2 \) & 18,500,124 & 7.88E-03 & 1.34E-03 \\ 
         & MgNO-\( H_2 \) & 18,500,124 & 7.92E-03 & 1.48E-03 \\ 
         & FNO-\( L_2 \) & 11,989,761 & 9.51E-03 & 7.99E-04 \\ 
         & FNO-\( H_1 \) & 11,989,761 & 9.34E-03 & 7.87E-04 \\ \midrule
        \multirow{4}{*}{K-S} & MgNO-\( L_2 \) & 18,500,124 & 2.28E-02 & 1.45E-03 \\
         & MgNO-\( H_1 \) & 18,500,124 & 2.40E-02 & 1.48E-03 \\ 
         & FNO-\( L_2 \) & 11,989,761 & 4.60E-02 & 9.07E-04 \\ 
         & FNO-\( H_1 \) & 11,989,761 & 4.08E-02 & 7.98E-04 \\ 
        \bottomrule
    \end{tabular}
\end{table}

\begin{figure}[h]
    \centering
    \includegraphics[width=0.8\textwidth]{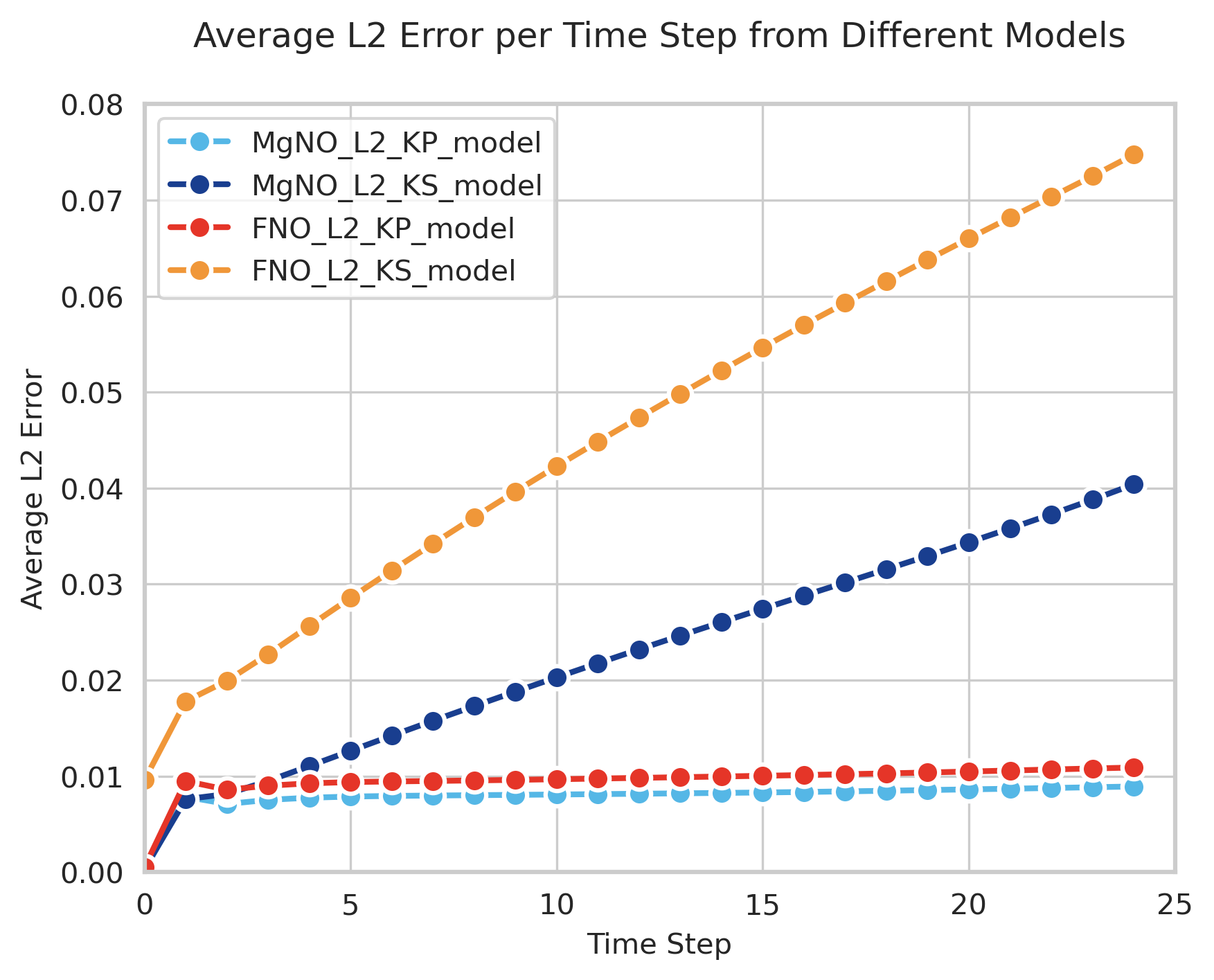}
    \caption{Errors of each model at time steps from 0 to 24}
    \label{fig:Modelplot}
\end{figure}

From Figure \ref{fig:Modelplot}, we analyze the average \(L_2\) error performance of the four models trained with the \(L_2\) function over time steps from 0 to 24.

In the K-P prediction task, both MgNO\({-L_2}\) and FNO\({-L_2 }\) models show robust and accurate prediction abilities. However, in the K-S prediction tasks, the FNO\({-L_2 }\) model exhibits a significant increase in error over time, indicating a potential accumulation of time-dependent error. The MgNO\({-L_2}\) model also shows an increase in error over time but at a much smaller rate than the FNO model, demonstrating better stability.

The MgNO and FNO models perform similarly in the K-P task. In the K-S task, since the presence of more complex behavior in saturation, MgNO outperforms FNO, being nearly twice as accurate, demonstrating its potential in handling complex spatiotemporal sequence predictions.

\subsection{Generalization Capability}

To assess long-term prediction performance, we extended the time series to 60 days, simulating longer trends. We trained models on data from time steps 0 to 24 and tested on 60-day sequences, where steps 0 to 24 were seen in the training dataset, and steps after 24 were not seen. 

\begin{figure}[h]
    \centering
        \includegraphics[width=\textwidth]{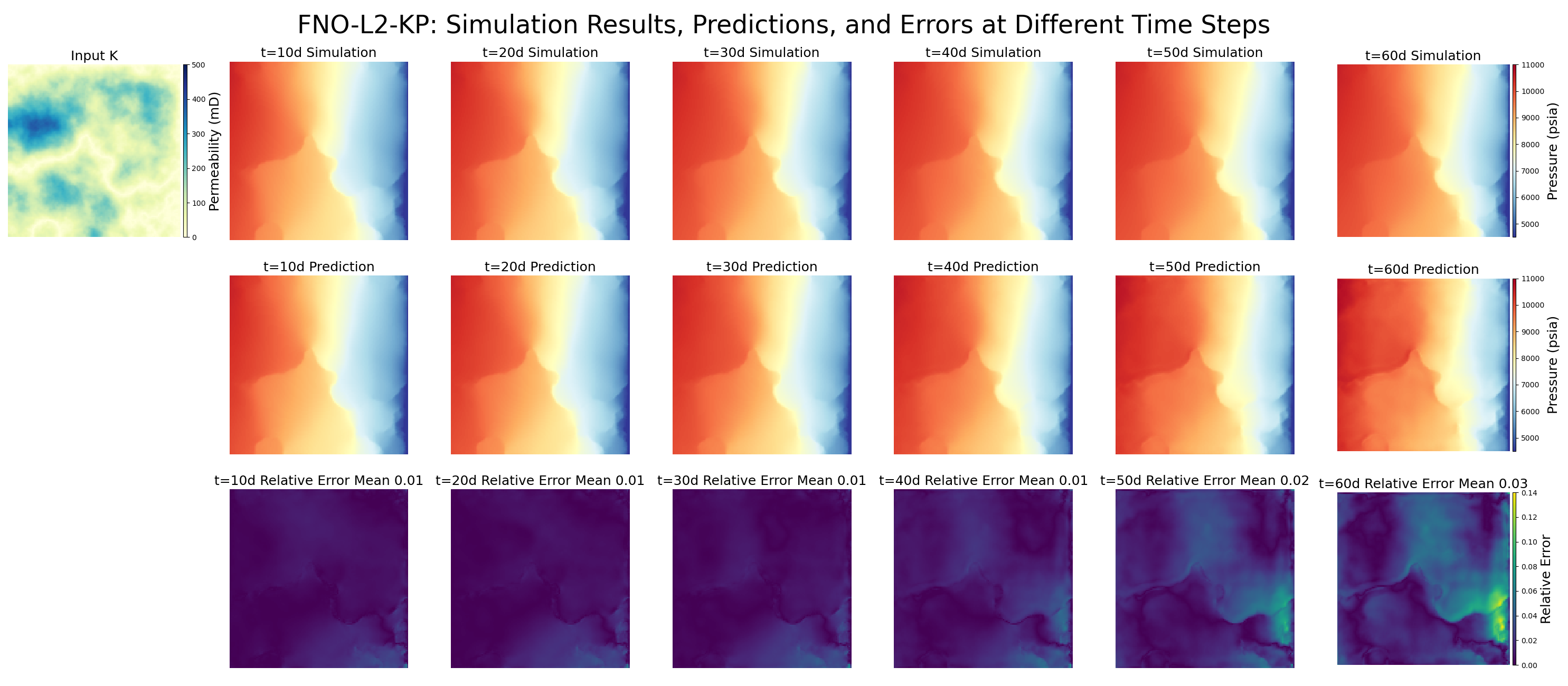}  
    
    \caption{K-P Model Results for FNO}
    \label{fig:KP_Model_Results_FNO}
\end{figure}

\begin{figure}[h]
        \centering
        \includegraphics[width=\textwidth]{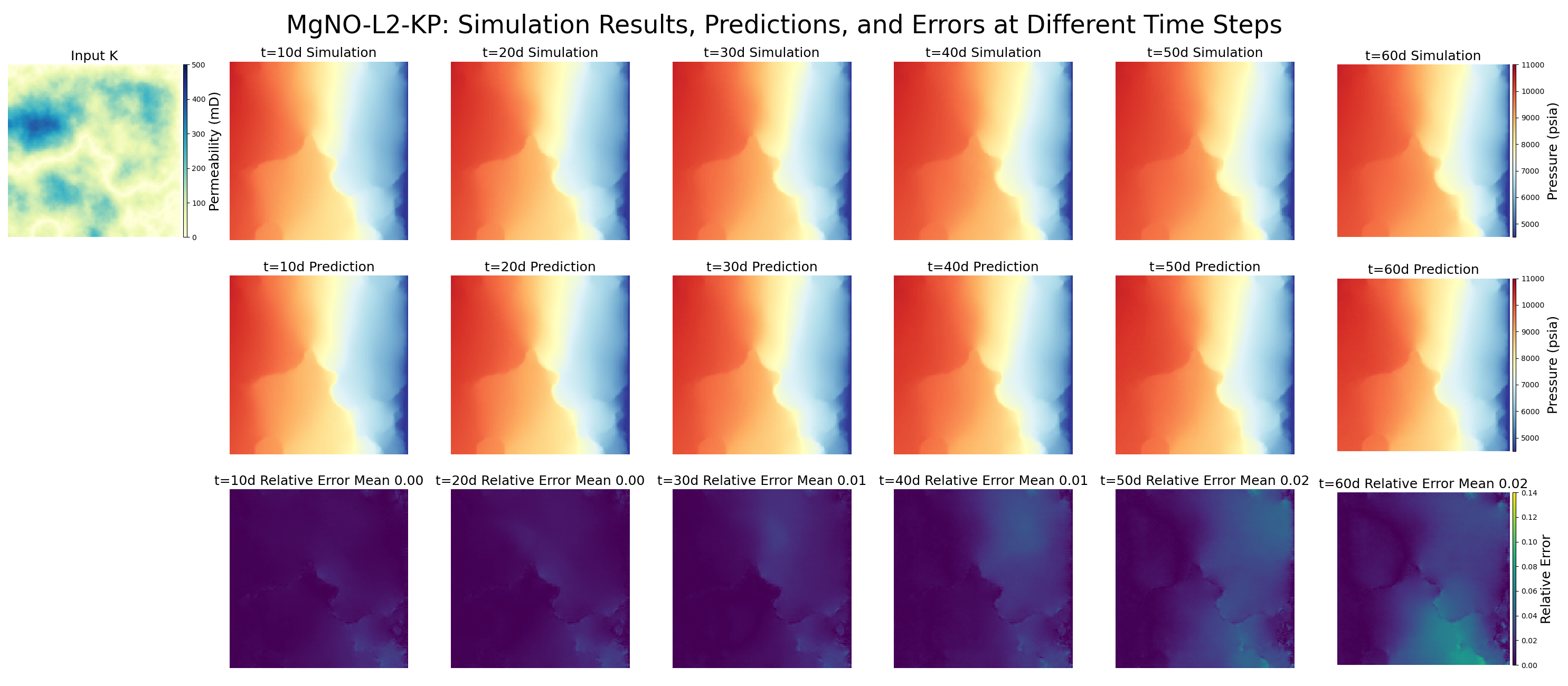}
   
    \caption{K-P Model Results for MgNO}
    \label{fig:KP_Model_Results_MgNO}
\end{figure}

\begin{figure}[h]
    \centering
  
        \includegraphics[width=\textwidth]{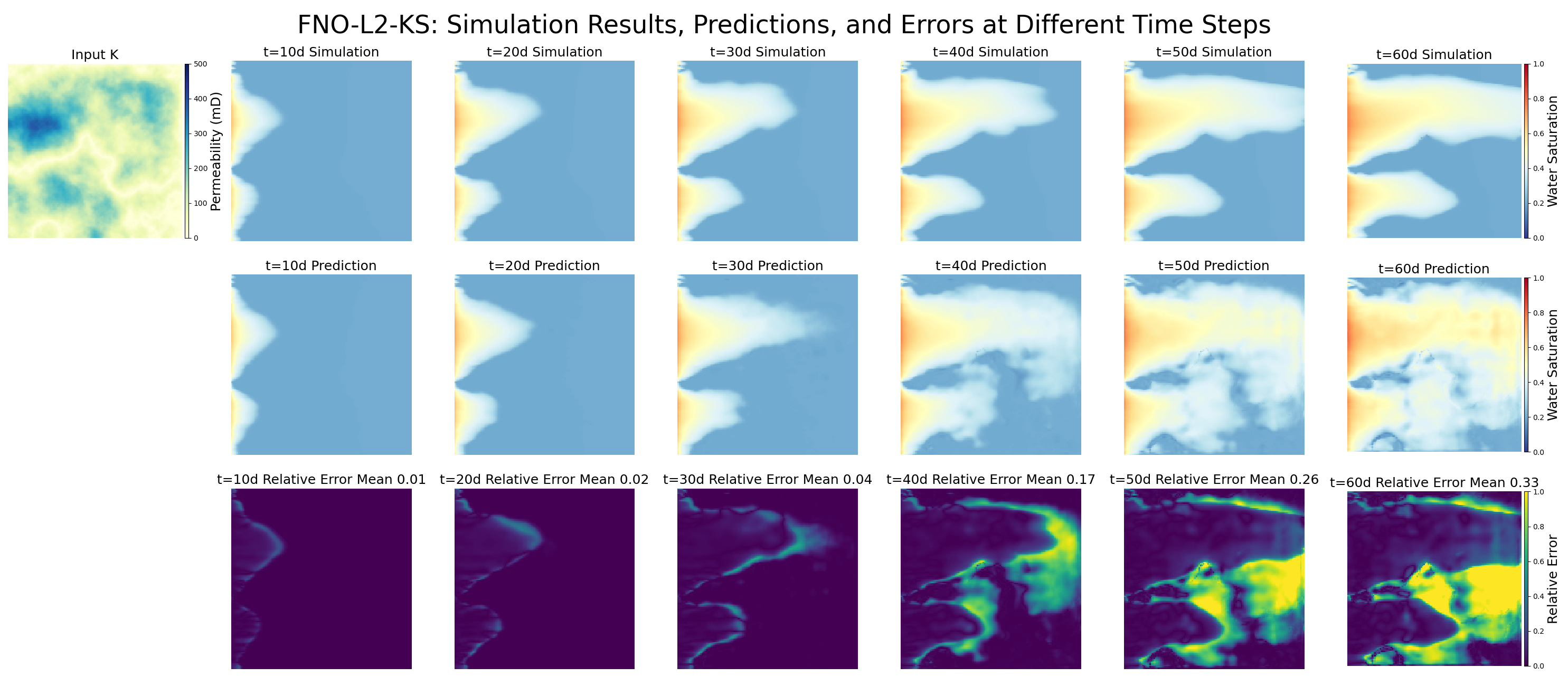}
    
    \caption{K-S Model Results for FNO}
    \label{fig:KS_Model_Results_FNO}
\end{figure}

\begin{figure}[H]
    \centering
        \centering
        \includegraphics[width=\textwidth]{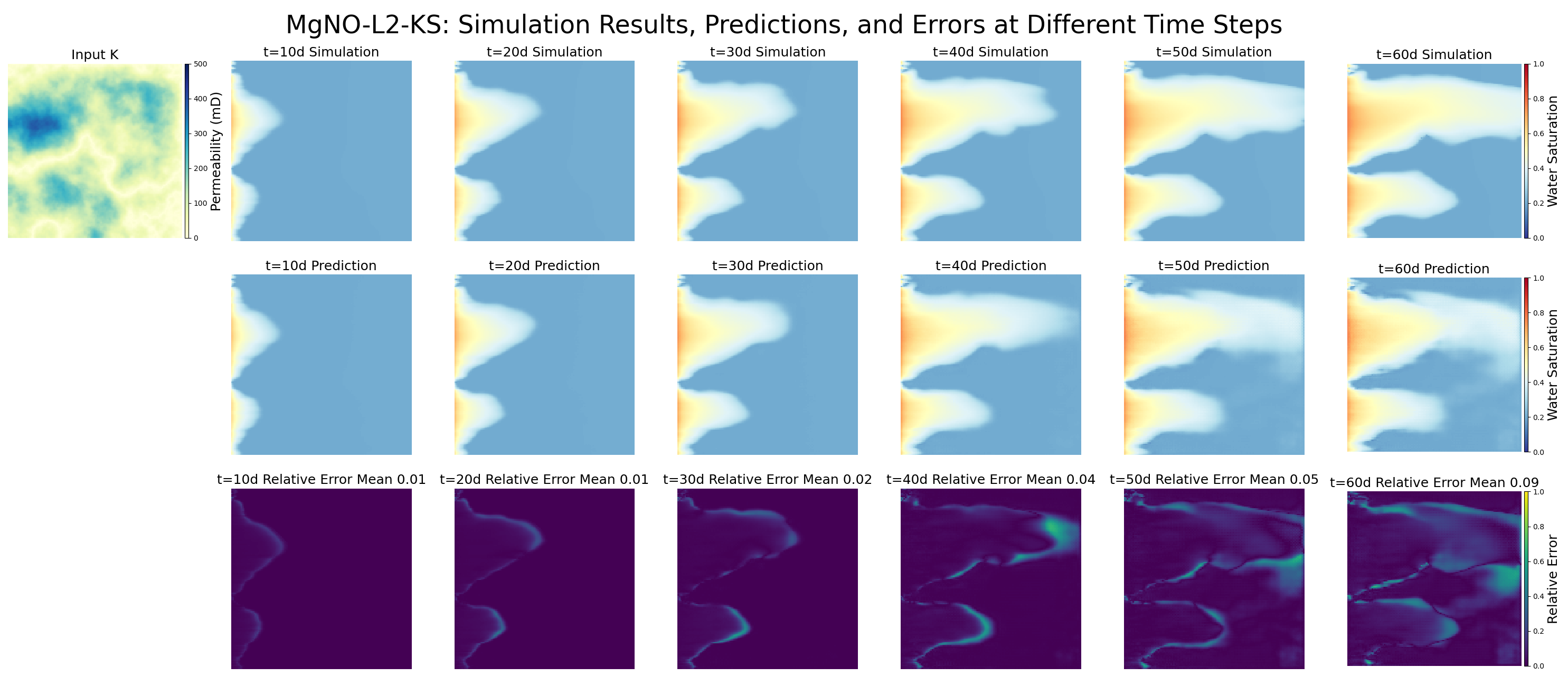}

    \caption{K-S Model Results for MgNO}
    \label{fig:KS_Model_Results_MgNO}
\end{figure}

Figures \ref{fig:KP_Model_Results_FNO} and \ref{fig:KP_Model_Results_MgNO} show K-P model results for FNO and MgNO, respectively. FNO models exhibit increasing error by day 60, while MgNO models maintain high accuracy, demonstrating superior generalization even on unseen time steps.
Similarly, Figures \ref{fig:KS_Model_Results_FNO} and \ref{fig:KS_Model_Results_MgNO} depict K-S model results. MgNO models maintain stability and accuracy, contrasting with FNO models that show significant error accumulation by day 60. MgNO's ability to generalize well on long-term predictions, including unseen time steps, highlights its potential in complex spatiotemporal sequence predictions in reservoir simulations.

\section{Conclusion}

The experimental results and model inference of this study indicate that while the FNO model performs reliably for shorter time series, it faces challenges of error accumulation in long-term prediction tasks. The MgNO model, on the other hand, demonstrates superior performance, particularly in long-term time series prediction tasks, exhibiting lower error accumulation, better generalization ability, and higher model stability.

The unique structural design of MgNO provides the capability to handle complex spatiotemporal data, effectively capturing and simulating dynamic changes over long time spans, which is especially significant in K-S long-term prediction tasks. Additionally, the performance of MgNO in terms of model scalability proves its structural potential for future applications in reservoir simulations and more complex scenarios.

This study explores the application of MgNO for reservoir simulation, showcasing the model's effectiveness and efficiency in solving complex PDEs. After training, MgNO can perform inference quickly, significantly reducing the time required for repeatedly solving PDEs. Its computational efficiency is over a hundred times faster than traditional numerical methods, making it particularly suitable for history matching and inverse problem-solving in reservoir simulations, providing faster feedback for real-time performance.

The study demonstrates the significant potential of the MgNO model in reservoir simulation applications. Future research will continue to optimize the structure and training process of MgNO to reduce application costs and expand its feasibility and impact in industrial practice. Additionally, the application of the MgNO model will be extended to other scientific computing tasks, such as atmospheric circulation simulation and ecosystem dynamics prediction, promoting the advancement of scientific frontiers.

%
%

\newpage
\appendix
\section{Appendix}
\subsection{Data Generation}
\subsubsection{Numerical Simulation Setup.}
We constructed a numerical simulation environment based on the oil-water two-phase flow model, using a three-dimensional grid with a size of 128×1×128 (xoz plane). The simulation domain is uniformly divided in all three spatial directions, with a step size of 10 meters. The entire simulation period is 24 days, with results output daily. Initial conditions are set with a porosity of 0.3 and a water saturation of 0.2. Boundary conditions include an injection well on the left boundary and a production well on the right boundary.
\subsubsection{Generation of Permeability.}
The distribution of absolute permeability \( K \) is generated through a Gaussian Random Field (GRF) with a mean of 0 and a covariance operator \( C = (-\Delta + 9I)^{-2} \), where \( \Delta \) is the Laplacian with zero Neumann boundary conditions. The generated permeability field \( K \) is converted to non-negative values and scaled such that \( K = 10 \times |K| \), thereby adjusting the flow velocity of the fluid in the medium.
\subsubsection{Numerical Method.}
A GRF is used to generate the Gaussian random field. This function utilizes the Discrete Cosine Transform (DCT) to perform the Karhunen-Loève expansion, calculating and returning samples of the random field. The specific implementation of the function is provided in Appendix 1.
\subsubsection{Software Configuration.}
The simulation was performed using the reservoir simulation software OpenCAEPoroX Version-0.5.0. The configuration details include fluid properties, rock physical data, relative permeability and capillary pressure curves, and PVT (Pressure-Volume-Temperature) relationships, ensuring the accuracy of the simulation results.
\subsubsection{Output Data Processing.}
The simulation outputs include key production data, such as daily oil production, daily water production, and wellhead pressure, as well as the pressure and water saturation distribution in the reservoir at each time step. To ensure the data is effectively used for training and validating machine learning models, we adjusted the output data to a specific format. The entire dataset comprises 2000 data sets, each including a specific permeability value (K) and the corresponding daily water saturation (S) and pressure (P) data from the initial time (day 0) to the end time (day 24).

To enhance data processing efficiency and ensure compatibility with machine learning frameworks, we transformed and segmented the 2000 simulation data sets as follows:
\begin{enumerate}
    \item \textbf{Data Transformation:} The raw simulation data is transformed into three main NumPy file formats:
    \begin{itemize}
        \item \textbf{K file:} Contains the permeability values (K) for each data set.
        \item \textbf{P file:} Stores the pressure values (P) for each data set, recording daily pressure from day 0 to day 24.
        \item \textbf{Sw file:} Records the water saturation values (Sw) corresponding to the P file.
    \end{itemize}

    \item \textbf{File Structure:}
    \begin{itemize}
        \item Each file is in .npy format, which is particularly suitable for storing multidimensional arrays and is easy to load and process in a Python environment.
        \item \textbf{K.npy:} Contains the permeability values (K) for each data set, with each K value repeated 25 times in the array to match the pressure and water saturation data for each time step. Thus, K.npy contains a 2000×25 array, with each row corresponding to the same permeability value repeated 25 times.
        \item \textbf{P.npy} and \textbf{Sw.npy} files both contain 2000×25 arrays, with each row representing a set of pressure or water saturation measurements at 25 time points.
    \end{itemize}

    \item \textbf{Data Loading and Usage:}
    \begin{itemize}
        \item These files can be directly loaded into any Python-supported data science or machine learning tool using the NumPy load function.
        \item This preprocessing method not only simplifies data management but also accelerates data reading speed, making subsequent data analysis and model training more efficient.
    \end{itemize}
\end{enumerate}

\subsection{Detailed Settings for Model and Experiments}

We trained the KP and KS models to predict the temporal variations of the pressure field and water saturation field in complex multiphase flow using the \( L_2 \) and \( H_1 \) loss functions following \cite{liu2024mitigating}, respectively. The training is expected to yield eight models: FNO-\( L_2 \)-KP, etc. The training settings are as follows: 500 epochs, batch size of 50. The 2000 dataset is divided into 1600 training sets and 400 validation sets. The parameters are set as weight decay: \( 1 \times 10^{-5} \), learning rate: \( 1 \times 10^{-4} \), scheduler: Adam. The hardware platform used for the experiments consists of 2 Huawei Kungpeng-920 @ 3.0GHz and 4 Nvidia A100 PCIe 40GB.
\end{document}